\documentclass[10pt]{article}
\usepackage{graphicx}
\usepackage{amsmath}

\def\Title#1{\begin{center} {\Large #1 } \end{center}}
\def\Author#1{\begin{center}{ \sc #1} \end{center}}
\def\Address#1{\begin{center}{ \it #1} \end{center}}

\newcommand\pubblock{\rightline{\begin{tabular}{l} Proceedings of the Fifth Annual LHCP\\ 
         \pubdate  \end{tabular}}}

\newenvironment{Abstract}{\begin{quotation} \begin{center} 
             \large ABSTRACT \end{center}\bigskip 
      \begin{center}\begin{large}}{\end{large}\end{center} \end{quotation}}

\newenvironment{Presented}{\begin{quotation} \begin{center} 
             PRESENTED AT\end{center}\bigskip 
      \begin{center}\begin{large}}{\end{large}\end{center} \end{quotation}}

\def\beq{\begin{equation}}
\def\eeq#1{\label{#1}\end{equation}}
\def\eeqn{\end{equation}}

\def\beqa{\begin{eqnarray}}
\def\eeqa#1{\label{#1}\end{eqnarray}}
\def\eeqan{\end{eqnarray}}

\let\bar=\overbar

\def\Dslash{\not{\hbox{\kern-4pt $D$}}}
\def\dslash{\not{\hbox{\kern-2pt $\del$}}}

\def\msb{{\bar{\ssstyle M \kern -1pt S}}}

\usepackage{color}

\newcommand\pubdate{\today}

\def\affiliation{
On behalf of the CMS Collaboration, \\
Department of Physics \\
University of Wisconsin -- Madison, Madison, WI 53706-1390, USA}

\begin{document}

\large
\begin{titlepage}
\pubblock

\vfill
\Title{  Multi-boson Measurements in CMS }
\vfill

\Author{ Kenneth Long }
\Address{\affiliation}
\vfill
\begin{Abstract}

Recent results from the CMS experiment on the 
production of multiple vector bosons in proton-proton collisions at the 
CERN LHC are presented. Measurements of WZ and ZZ production with 
fully leptonic decays, Z$(\nu\nu)\gamma$, and 
WV (V $=$ W,Z $\rightarrow q\bar{q}$) production at 8 and 13 TeV are discussed. 
Selected cross section measurements, unfolded differential measurements, and 
limits on anomalous triple gauge couplings
are presented and compared with theoretical predictions.

\end{Abstract}
\vfill

\begin{Presented}
The Fifth Annual Conference\\
 on Large Hadron Collider Physics \\
Shanghai Jiao Tong University, Shanghai, China\\ 
May 15-20, 2017
\end{Presented}
\vfill
\end{titlepage}
\def\thefootnote{\fnsymbol{footnote}}
\setcounter{footnote}{0}
%

\normalsize 


\section{Introduction}

Measurements of the production of multiple vector bosons in proton-proton collisions 
at the CERN LHC is an important test of the electroweak sector of
the Standard Model (SM). Such final states provide a natural probe of the 
vector boson self-couplings, which arise due to the non-Abelian nature of the
electroweak gauge group and are exactly predicted by the SM. Deviations from these
predictions could therefore be an indication of new physics. In addition,
due to recent theoretical developments, the production cross sections 
of all processes producing vector boson pairs in proton-proton collisions are
known at next-to-next-to-leading order (NNLO) in QCD -- many differentially. 
When the perturbative QCD corrections are significant with respect to the experimental
uncertainties, measurements can directly validate the higher-order perturbative QCD
calculations.

This report summarizes recent measurements from the CMS collaboration using
data collected by the CMS detector at $\sqrt{s} =$ 8 and 13 TeV. A brief summary
of measurements for a variety of vector boson pairs and decays is presented, 
each offering a unique probe of the SM and possible new physics. Full details of 
the analyses can be found in the references.

\section{Z$(\nu\nu)\gamma$ Production at 13 TeV}
In the SM, proton-proton collisions producing a Z boson in association with a photon,
with the Z decaying to a pair of neutrinos, can only proceed through
initial-state radiation of a photon from the incoming quarks.
Other production mechanisms could arise through trilinear gauge boson 
self-interaction couplings (TGCs) which would be a sign of new physics. 
Measurements of this channel for high photon transverse energy are therefore
sensitive to new physics, and are also important as background measurements to
dark matter searches in the mono-photon channel.

The $pp \rightarrow \mathrm{Z}\gamma \rightarrow \nu\bar{\nu}\gamma$ fiducial cross section 
has been measured for transverse energy $E_{T}^{\gamma} > 175$ GeV and photon pseudorapidity
$\lvert\eta^{\gamma}\rvert < $ 1.44 using 2.3 fb$^{-1}$ of data collected by the CMS experiment in 2015
\cite{CMS-PAS-SMP-16-004}.
The analysis selects events with a well-identified photon with transverse energy $E_{T}^{\gamma} > 175$ GeV
and missing transverse momentum $p_{T}^{\mathrm{miss}} > 170$ GeV. 
The two are required to
have large angular separation and to be separated from all jets in the event, as
the missing transverse momentum in $\gamma$+jet background events is typically aligned 
with mismeasured jets.
The primary backgrounds to the signal process are experimental in nature, including 
spurious signals in the electromagnetic calorimeter (ECAL), beam halo, and cosmic rays. 
These contributions all give rise to signals mimicking photons in the detector, 
with missing transverse momentum arising due to the 
condition that the transverse momentum of the event sum to zero. The contribution of
these backgrounds is estimated
by exploiting their characteristic shape and timing in the ECAL. Other backgrounds 
enter the selection due to misidentification or objects escaping the detector acceptance,
including $\gamma+$jets, W$\rightarrow \ell\nu$, $Z(\rightarrow\ell\ell)\gamma$ and 
W$(\rightarrow\ell\ell)\gamma$. They are estimated from simulated samples, with the exception
of W decays to electrons and QCD multijet backgrounds which are estimated from data.

The measured fiducial cross section of
\begin{equation}
  \sigma_{\mathrm{fid}}(pp \rightarrow Z\gamma \rightarrow \nu\bar{\nu}\gamma) = 66.5 \pm 13.6 \, \mathrm{(stat)} \pm 14.3 \, 
        \mathrm{(syst)} \pm 2.2 \, \mathrm{(lumi)} \,\mathrm{fb}
\end{equation}
is found to be in agreement with the SM prediction of $65.5 \pm 3.3$ fb, computed at
NNLO with \textsc{MATRIX}. \cite{Grazzini:2015nwa}

\section{W$^{\pm}$Z Production with Leptonic Decays at 8 and 13 TeV}

Measurements of WZ production with fully leptonic decays probe the charged
gauge interactions of the SM. The relatively clean three-lepton final state
is balanced by a modestly high cross section, allowing a high statistics measurement 
with controlled systematics. The analysis selects events with three leptons and 
missing transverse momentum $p_{T}^{\mathrm{miss}} > 30$ GeV. The condition on the
mass of the three lepton system
$m_{3\ell} > $ 100 GeV is required to reject Z$\gamma$ events where a photon
radiated from the leptonic Z decay is misidentified as an electron. Events with
a b-tagged jet are rejected to reduce background yields from $t\bar{t}$.

\begin{figure}[htb]
  \centering
    \includegraphics[height=2.4in]{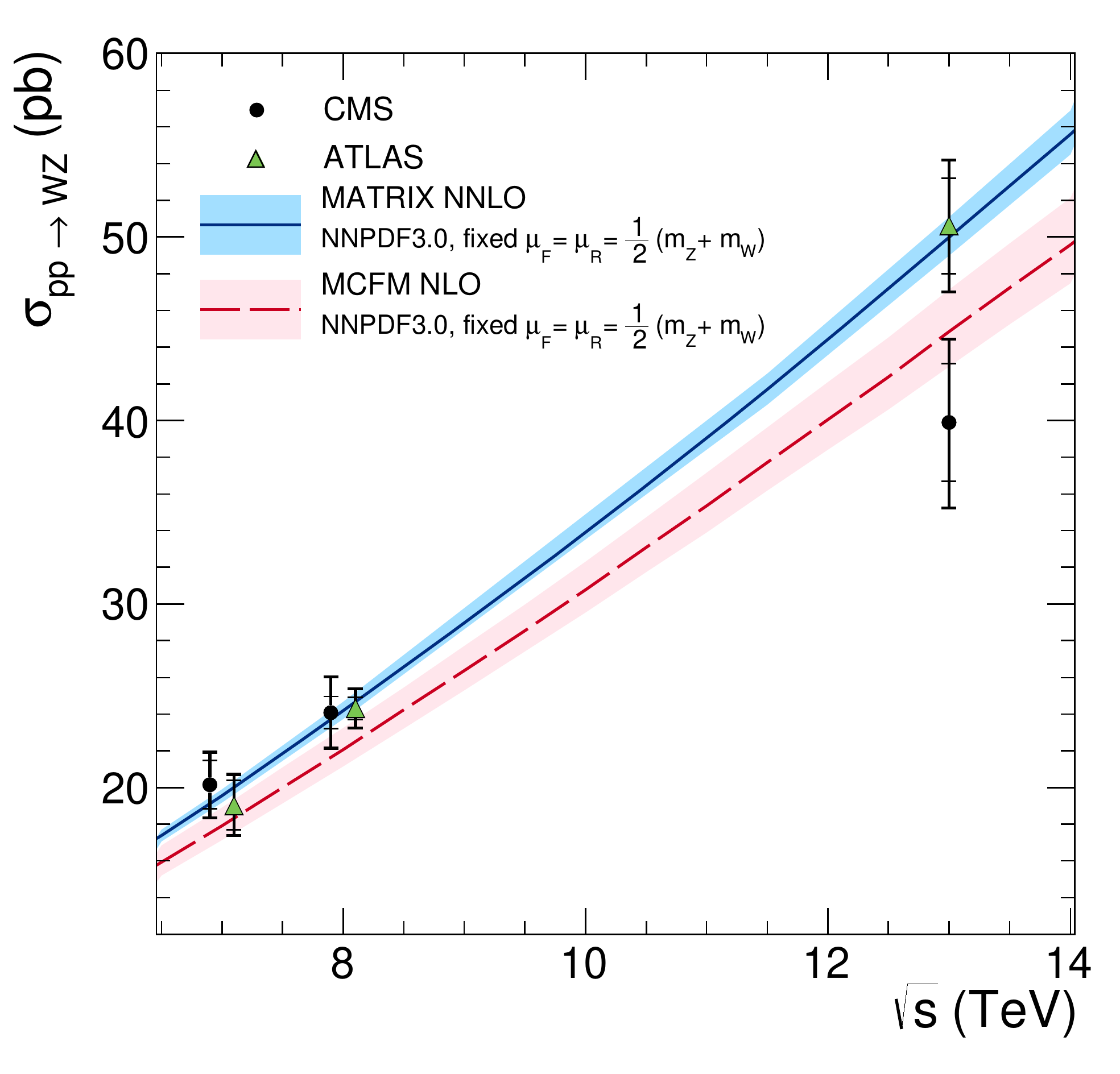}
    \includegraphics[height=2.4in]{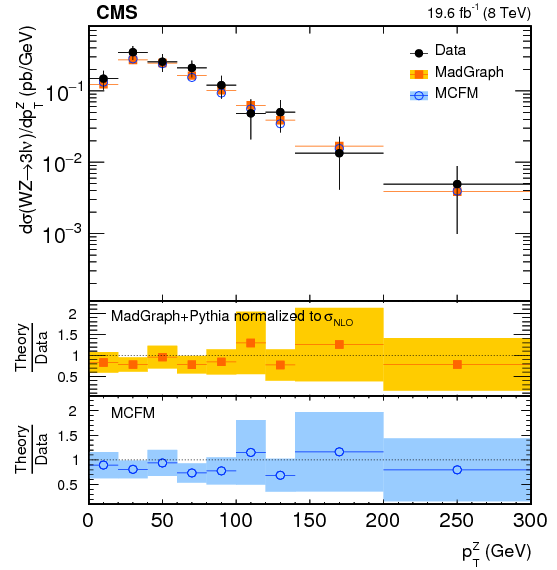}
    \caption{ (Left) The total $pp \rightarrow $WZ cross section
      as a function of $\sqrt{s}$ measured by the CMS and 
      ATLAS experiments compared to the predictions of \textsc{MCFM} 7.0 and \textsc{MATRIX} v1.0.0\_beta4. 
      The CMS 13 TeV cross section are calculated for the Z boson mass window $60 - 120$ GeV. 
      The CMS 7 and 8  TeV  cross sections are calculated for the Z boson mass window $71 - 111$ GeV.
      ATLAS measurements are performed with the Z boson mass window $66 - 116$  GeV.
      (Right) Differential cross section at $\sqrt{s} = 8$ TeV as a function
      of the Z boson transverse momentum, compared to
      the fixed-order prediction from \textsc{MCFM} 6.3 and the prediction from
      MadGraph5.1+\textsc{PYTHIA}6.4, using MLM merging of tree-level contributions 
      for up to two additional partons~\cite{Khachatryan:2016poo}.
      }
  \label{fig:WZfigs}
\end{figure}

Backgrounds for this measurement are categorized into processes producing at least
three prompt isolated leptons ($\ell = e, \mu$), including ZZ, triboson, and 
$t\bar{t}\gamma$, and those processes where nonprompt
leptons from hadrons decaying to leptons inside jets or jets misidentified as isolated
leptons pass the signal selection, predominantly $t\bar{t}$ and Drell-Yan. 
The nonprompt background is evaluated with a data-driven approach using 
control regions of events passing the full analysis selection,
with the exception that one, two, or three leptons pass relaxed identification 
and isolation requirements but fail the more stringent requirements applied to signal events.
These events are extrapolated into the signal region using per-lepton 
``tight to loose'' transfer factors
calculated from a sample of dijet events. Prompt backgrounds are evaluated using 
simulated samples. The Z$\gamma$ process is also estimated with simulation.

The production cross section has been measured by the CMS experiment at 7, 8, 
\cite{Khachatryan:2016poo}
and 13 TeV
\cite{Khachatryan:2016tgp}.
These measurements and their agreement with the SM
predictions, computed at NNLO in QCD with \textsc{MATRIX}~\cite{Grazzini:2016swo} and 
next-to-leading order (NLO) in 
QCD with \textsc{MCFM}~\cite{Campbell:2011bn}
are summarized in Fig.~\ref{fig:WZfigs}, which also presents ATLAS results for comparison. 
Good agreement with the SM predictions
is observed. Some tension is seen in the 13 TeV measurement with 2.3 fb$^{-1}$
and the NNLO prediction, but we note that this measurement is statistically
limited and await a measurement with an increased 13 TeV dataset.

Differential measurements of WZ production, unfolded using the 
iterative d'Agostini method~\cite{DAgostini:1994fjx}, are presented using 19.6 fb$^{-1}$
collected by the CMS experiment in 2012 at 8 TeV. These measurements allow for
comparisons of theoretical predictions in distributions sensitive 
to higher-order corrections, such as the Z boson transverse momentum, which is
shown in Fig.~\ref{fig:WZfigs}. Limits on anomalous quartic gauge couplings are
also calculated at 8 TeV, details of which can be found in the reference.

\section{ZZ Production with Leptonic Decays at 8 and 13 TeV}

In spite of its low cross section, the ZZ $\rightarrow 4\ell$ process 
is favorable experimentally due to its clean and fully reconstructed 
four-lepton final state. It provides a probe of the neutral gauge sector 
of the SM, and its role as the primary background to the SM Higgs
boson in the four-lepton channel makes it an important
process to understand.

\begin{figure}[htb]
  \centering
    \includegraphics[height=2.4in]{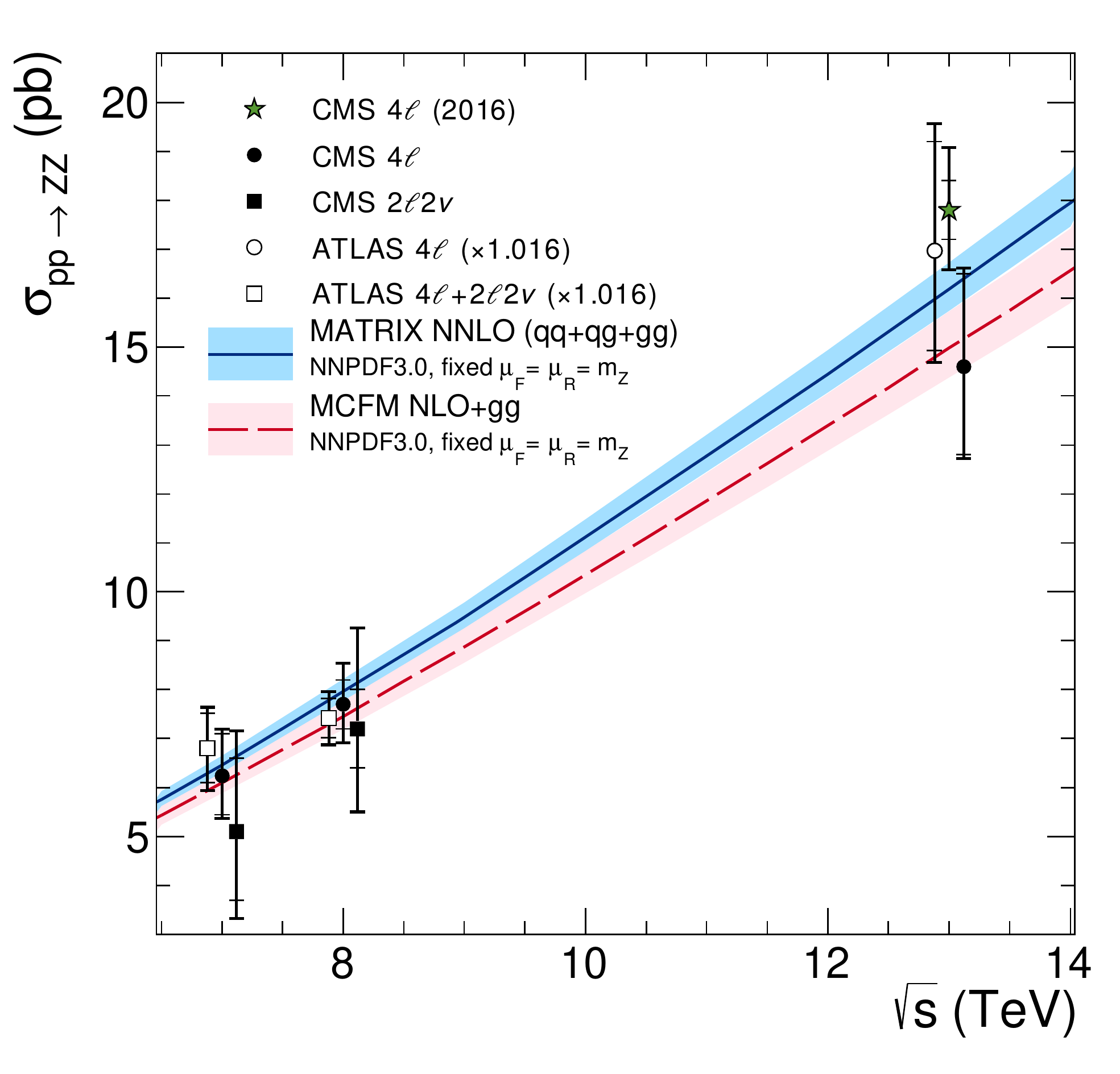}
    \includegraphics[height=2.3in]{ZZ_13TeV_mZZ_unfolded.pdf}
    \caption{ (Left) The total $pp \rightarrow$ ZZ cross section
      as a function of $\sqrt{s}$ measured by the CMS and 
      ATLAS experiments compared to the predictions of \textsc{MCFM} 7.0 and \textsc{MATRIX} v1.0.0\_beta4. 
      Correction factors are applied to the ATLAS measurements to account
      for the different Z boson mass windows used for the two measurements.
      (Right) Normalized differential cross section as a function
      of the four-lepton mass compared to the predictions from
      MadGraph5\_aMC@NLO and \textsc{POWHEG} v2, both showered using \textsc{PYTHIA} 8~\cite{CMS:2017ruh}.
      }
  \label{fig:ZZinclusive}
\end{figure}

Measurements of the ZZ production cross section, inclusive differential distributions,
and differential distributions of ZZ production associated with jets
were performed using 35.9 fb$^{-1}$ of 13 TeV data collected by the CMS experiment in 2016
\cite{CMS:2017ruh}\cite{CMS-PAS-SMP-16-019}
and 19.7 fb$^{-1}$ of data collected at 8 TeV in 2012~\cite{CMS-PAS-SMP-15-012}. 
The analysis selects four leptons forming two 
Z boson candidates with
mass in the range $60 < m_{\ell^{+}\ell^{-}} < 120$ GeV. The very low background
in this channel permits the use of loose identification and isolation criteria. 
Backgrounds with four prompt leptons such as triboson and $t\bar{t}V$, 
which have very
low production cross sections, are evaluated using simulated samples. The 
contribution from nonprompt leptons is estimated using the same technique
described for WZ in the previous section. However, the ``tight to loose'' transfer 
factors are derived using a sample of Drell-Yan events, with associated 
jets as the loose and tight lepton probe.

\begin{figure}[htb]
  \centering
    \includegraphics[height=2.4in]{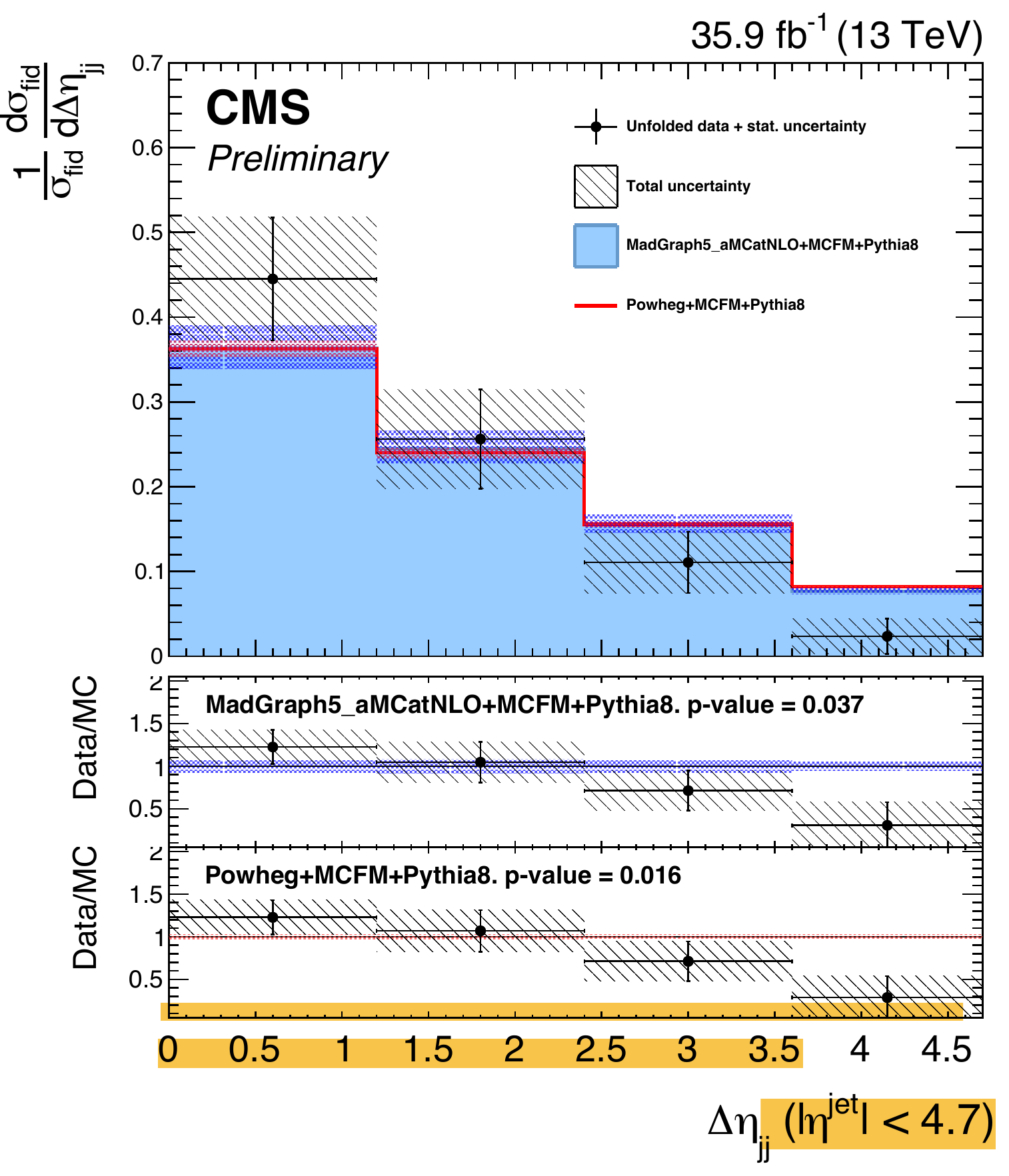}
    \includegraphics[height=2.4in]{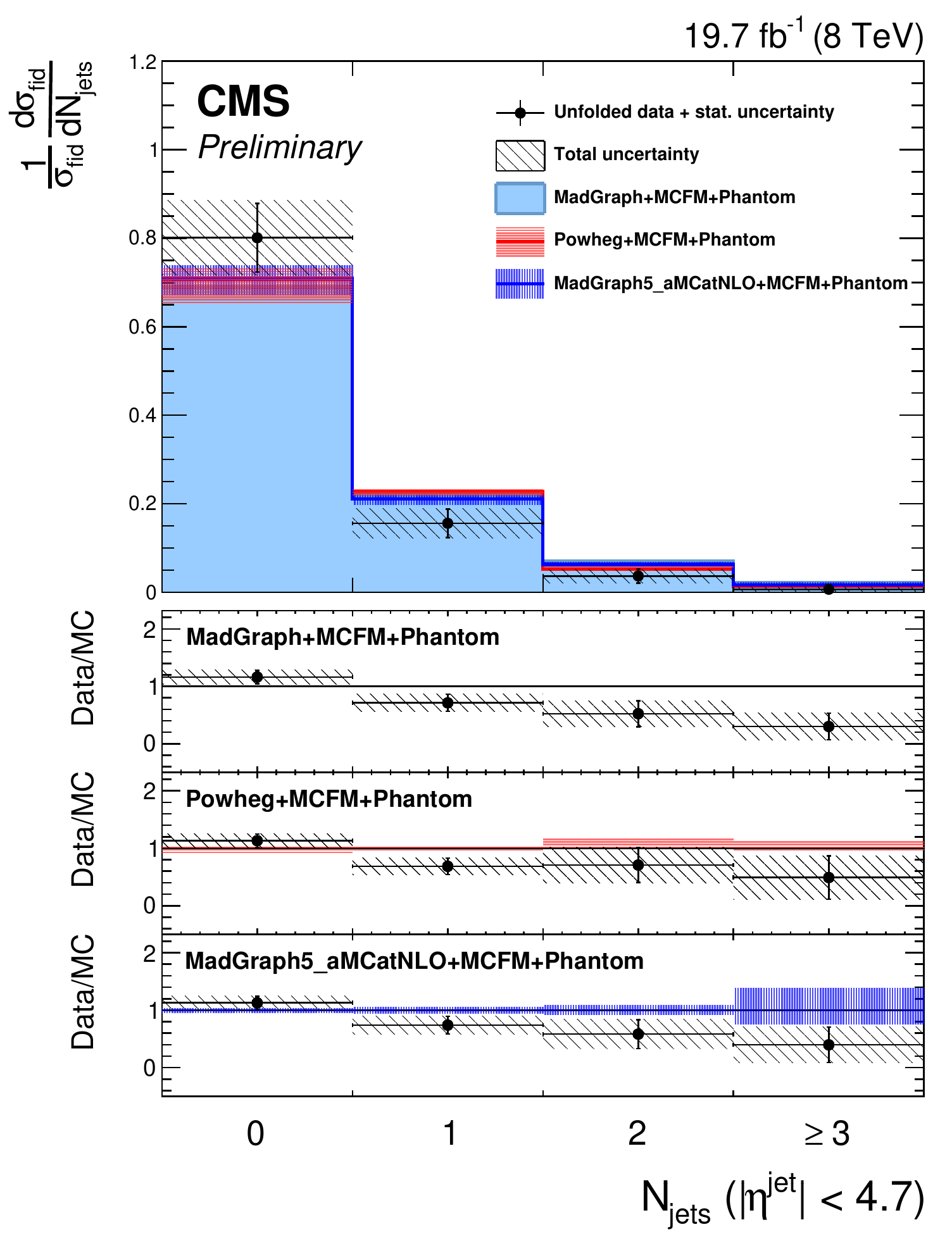}
  \caption{(Left) Normalized differential cross section as a function of the
          leading jet pseudorapidity separation in ZZ events with at least
          two anti-$k_{T}$ $R=0.4$ jets at 13 TeV~\cite{CMS-PAS-SMP-16-019}. 
          (Right) Normalized differential cross section at 8 TeV
          as a function the number of anti-$k_{T}$ $R=0.5$ jets in the 
          event~\cite{CMS-PAS-SMP-15-012}.}
  \label{fig:ZZjets}
\end{figure}
The total cross section measurements and their agreement with the SM
prediction, computed at NNLO in QCD with \textsc{MATRIX}~\cite{Cascioli:2014yka}\cite{Grazzini:2015hta} 
and NLO in QCD with \textsc{MCFM}
are summarized in Fig.~\ref{fig:ZZinclusive}, which also presents ATLAS results for comparison. 
Differential cross section measurements have also been performed for a variety of inclusive 
and jet-dependent distributions, unfolded using the iterative d'Agostini method. 
The four-lepton mass, which includes contributions 
from resonant ZZ production, Z$\rightarrow 4\ell$ production, and $H \rightarrow 4\ell$ production,
is also presented in Fig.~\ref{fig:ZZinclusive}.
For this differential measurement the mass constraint of the Z boson pairs is relaxed to
$4 < m_{\ell^{+}\ell^{-}} < 120$ GeV, $40 < m_{\ell'^{+}\ell'^{-}} < 120$ GeV. 
Results are in good agreement with the SM predictions.

Differential measurements of ZZ boson production associated with jets are
a direct test of higher-order calculations. The normalized differential 
ZZ production cross section by number of anti-$k_{t}$ $R=0.5$ jets 
at 8 TeV is shown in Fig.~\ref{fig:ZZjets}. Also shown is the differential
cross section for the pseudorapidity separation of the two leading
anti-$k_{t}$ $R=0.4$ jets at 13 TeV for events with at least two jets.
Understanding this distribution is critical for extracting information
about vector boson scattering and vector boson quartic interactions.
Limits on anomalous quartic gauge couplings have also been calculated at 13 TeV, 
details of which can be found in the reference.

\section{WV Production with Semileptonic Decays}

The large branching fraction 
of vector bosons to quark anti-quark pairs offsets
the challenge of reconstructing the vector boson from
hadronic decay products. Diboson pairs with semileptonic decays
are therefore attractive due to the balance of leptonic signature, 
distinguishable above the large QCD multijet background at the LHC,
and the favorable branching fraction of the hadronic decay.
In particular, the higher production cross section
allows sensitivity to tails of distributions most sensitive to new physics
effects. Boosted object techniques allow further discrimination of
vector boson-like objects from background, and are particularly applicable
in the new physics regime.

\begin{figure}[htb]
  \centering
    \includegraphics[height=2.15in]{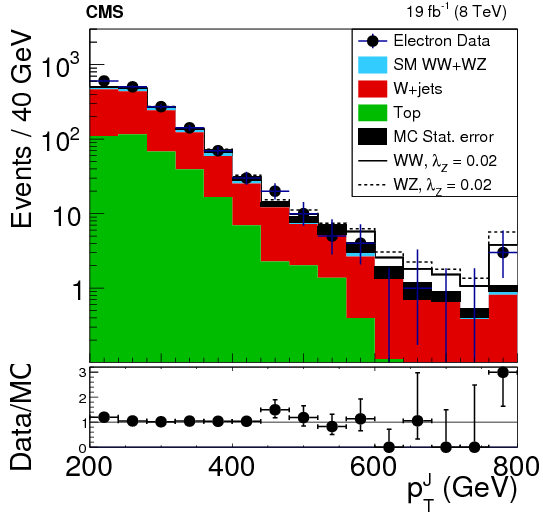}
    \includegraphics[height=2.15in]{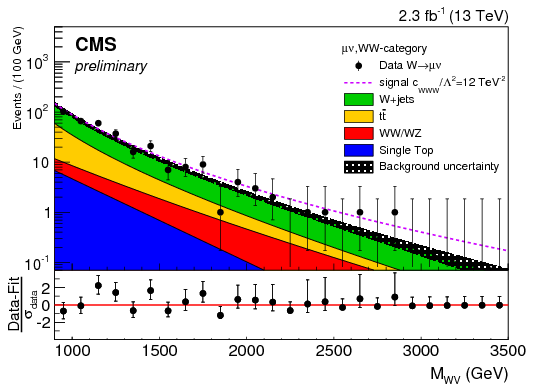}
  \caption{(Left) Transferse momentum of vector boson-tagged jet, used for
    aTGC limit extraction at 8 TeV~\cite{Sirunyan:2017bey}. (Right) Diboson mass, used for limit
    extraction in 13 TeV analysis~\cite{CMS:2016djf}. }
  \label{fig:WV}
\end{figure}

Limits on charged aTGC parameters, using the effective Lagrangian formalism 
of~\cite{Hagiwara:1993ck}, are calculated using 19.6 fb$^{-1}$ of 8 TeV 
data~\cite{Sirunyan:2017bey}
and 2.3 fb$^{-1}$ of 13 TeV data collected in 2015~\cite{CMS:2016djf}.
The analyses selects a high $p_{T}$ lepton and large missing transverse momentum
associated with the leptonically decaying W boson. An anti-$k_{T}$
(Cambridge-Aachen) ``fat jet''  is selected at 13 (8) TeV with $p_{T} > 200$ GeV 
and tagged as a vector boson candidate using
subtructure techniques including prunning~\cite{Ellis:2009me} and 
N-subjettiness~\cite{Thaler:2010tr}. 
Events with additional b-tagged jets are 
rejected to reduce $t\bar{t}$ background contributions. The analysis uses event categorisations
based on the jet mass to tag events as WZ-like or WW-like, but cannot fully 
distinguish the two states.

Limits are extracted via a fit to the WV invariant mass (transverse momentum of the vector boson-tagged jet) 
distribution at 13 (8) TeV.
These distributions are shown in Fig.~\ref{fig:WV}.
The 95\% CL limits from the 8 TeV analysis,
\begin{equation}
  -0.011 < \lambda_{Z} < 0.011, \quad -0.044 < \Delta\kappa_{\gamma} < 0.063, \quad -0.0087 < \Delta g^{Z} < 0.024,
\end{equation}
are the most stringent to date on $\Delta\kappa_\gamma$ and $\Delta g^{Z}$.

\section{Conclusions}
\begin{figure}[htb]
  \centering
    \includegraphics[height=2.35in]{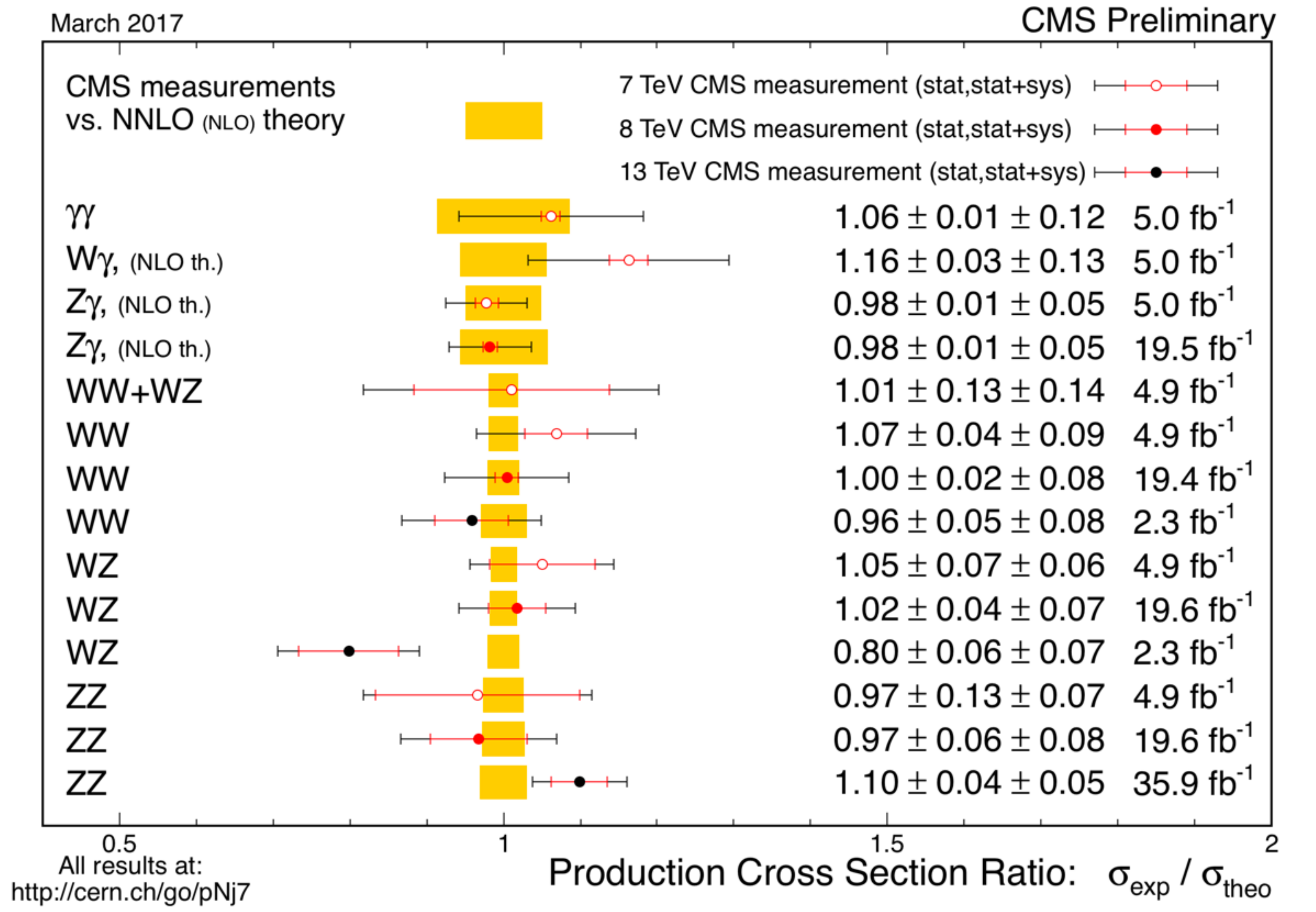}
  \caption{ Summary of diboson total production cross sections measured by the CMS experiment. }
  \label{fig:xsecs}
\end{figure}

We have summarized recent diboson measurements from the CMS experiment using
data collected at 8 TeV and 13 TeV. The current status of diboson cross sections 
measured by the CMS experiment is shown in Fig.~\ref{fig:xsecs}. We note that 
almost all results are compared to theoretical predictions at NNLO, where theoretical 
errors are typically below experimental ones, which 
presents an exciting challenge
for future measurements. 

\bibliographystyle{JHEP}
\bibliography{bibliography}
\end{document}